\documentstyle[12pt,epsf]{article}
\renewcommand{\bar}[1]{\overline{#1}}

\begin{document}

\bigskip\bigskip
\begin{center}

{\large \bf Role of T-odd functions\\
in high energy hadronic collisions}

\end{center}
\vspace{24pt}

\begin{center}
 {\bf Elvio Di Salvo\\}

 {Dipartimento di Fisica and I.N.F.N. - Sez. Genova, Via Dodecaneso, 33 \\-
 16146 Genova, Italy\\}
\end{center}

\vspace{24pt}
\begin{center} {\large \bf Abstract}

I propose a simple model for predicting the enegy behavior of T-odd, chiral odd 
function $h_1^{\perp}$. Furthermore I illustrate a method for extracting 
$h_1^{\perp}$ and the transversity function from Drell-Yan. The method may be 
applied also to other reactions.
\end{center}

\vspace{24pt}

\centerline{PACS numbers: 13.85.Qk, 13.88.+e}

\newpage

\section{Introduction}

The T-odd functions[1-3] have become important in the last ten years, since 
when high energy physicists realized that such functions could be used as 
{\it polarimeters} for extracting chiral odd functions, especially 
transversity[4-6]. Here I consider the T-odd, chiral odd function 
$h_1^{\perp}$[7] of a quark inside the proton. In particular I propose a simple 
model, which allows to predict the behavior of this function at varying proton 
momentum. Moreover I am concerned with asymmetries relative to unpolarized and 
singly polarized Drell-Yan (DY), {\it i. e.},
\begin{equation}
p p \to \mu^+ \mu^- X. \label{r1}
\end{equation}
I show how to extract $h_1^{\perp}$ from this reaction
and I suggest an alternative method for determining transversity.

\section{General formulae}

The single transverse spin asymmetry for reaction (\ref{r1}) is defined as
\begin{equation}
A = \frac{d\sigma_{\uparrow}-d\sigma_{\downarrow}}
{d\sigma_{\uparrow}+d\sigma_{\downarrow}},\label{asy0}
\end{equation}
where $d\sigma_{{\uparrow}(\downarrow)}$ refer to cross sections with 
opposite polarizations of one of the proton beams. In one-photon approximation, 
\begin{eqnarray}
d\sigma_{\uparrow}-d\sigma_{\downarrow} &\propto& d\Gamma 
L^{\mu\nu}H^a_{\mu\nu}, \label{diff0}
\\
L_{\mu\nu} &=& k_{\mu}k'_{\nu} +k'_{\mu}k_{\nu}-g_{\mu\nu}k\cdot k',\label{lt}
\\
H^a_{\mu\nu} &=& \int d^2p_{1\perp} Tr\left[\gamma_{\mu}\Phi_{\chi.o.}
(x_1,{\bf p}_{1\perp})\gamma_{\nu}{\bar \Phi}_{\chi.o.}(x_2,{\bf 
p}_{2\perp}) + (1\leftrightarrow 2)\right].\label{tens}
\end{eqnarray}
Here $d\Gamma$ is the phase space element.
$k$ and $k'$ are the four-momenta of the muons.
$x_1$ and $x_2$ are the longitudinal fractional momenta of the annihilating
quark and antiquark, ${\bf p}_{1\perp}$ and ${\bf p}_{2\perp}$ their 
transverse momenta with respect to the initial beams and $\Phi_{\chi.o.}$ and 
${\bar \Phi}_{\chi.o.}$ the chiral odd components of their correlation 
matrices. The index 1 in $x$ and ${\bf p}_{\perp}$ refers to the transversely
polarized proton, the index 2 to the unpolarized one. ${\bf p}_{2\perp}$ is 
chosen in such a way that the transverse momentum of the muon pair with 
respect to the proton beam in the laboratory frame, {\it i. e.},
\begin{equation}
{\bf Q}_{\perp} = {\bf p}_{1\perp}+{\bf p}_{2\perp},\label{cons}
\end{equation}
is kept fixed. Lastly the sum over flavors has been omitted. 
 
\section{Parametrization of the T-odd correlation matrix}

In the laboratory frame, at sufficiently high energies, the chiral odd 
component of the correlation matrix of the transversely polarized proton can be 
parametrized as[7]
\begin{equation}
\Phi_{\chi.o.} = \frac{1}{4}x_1{\cal P}\gamma_5\left\{[\rlap/S,\rlap/n_+]
h_{1T} + \frac{1}{\mu}[\rlap/r_{\perp},\rlap/n_+]h^{\perp}_1\right\}. 
\label{chir}
\end{equation}
Here $h_{1T}$ is the transverse momentum dependent transversity distribution, 
while $h^{\perp}_1$ will be illustrated in a moment. Moreover
\begin{eqnarray}
r_{\perp} &=& p_{1a}S - p_{1b}n_a \equiv (0,-p_{1b},p_{1a},0),
\\
p_{1a} &=& {\bf p}_{1\perp}\cdot {\bf S}\times {\bf n}, ~~~~ \ ~~~~~ \  p_{1b} 
= {\bf p}_{1\perp}\cdot{\bf S},
\\ 
\ n_+ &\equiv& (1, {\bf n}), ~~~~ \ ~~~~~ \ ~~~~ \  n_a \equiv 
(0, {\bf S}\times {\bf n}).
\end{eqnarray}
${\cal P}{\bf n}$ and $S\equiv (0,{\bf S})$ are respectively the momentum and
the Pauli-Lubanski four-vector of proton 1, ${\bf S}$ and ${\bf n}$ being unit 
vectors such that ${\bf S}\cdot {\bf n}$ = 0. Lastly $\mu$ is an 
undetermined mass scale, which was set equal to the proton mass by various 
authors[7-9]; as I shall show, this is not the most suitable choice. 

The second term of parametrization (\ref{chir}) is T-odd and gives 
a nonvanishing contribution also when the proton is unpolarized. In this case, 
given a unit vector ${\bf s}$ not parallel to {\bf n}, the density of quarks 
whose spin component along ${\bf s}$ is positive, minus the density of quarks 
for which this spin component is negative, amounts to
\begin{equation}
\delta q_{\perp} = -\frac{r_{\perp}\cdot s_0}{\mu}h_1^{\perp}, 
\label{polr}
\end{equation}
where $s_0\equiv (0,{\bf s})$. Eq. (\ref{polr}) is a consequence of eq. 
(\ref{chir}) for an unpolarized proton. The two equations exhibit the meaning 
of the function $h_1^{\perp}$: in an unpolarized proton, a quark with nonzero 
transverse momentum is polarized perpendicularly to its momentum and to the 
proton momentum, in agreement with parity conservation.

\section{A model for T-odd functions}

A proton may be viewed as 
a bound state of the active quark with a set $X$ of spectator partons.
In order to take into account coherence effects, I project the bound 
state onto scattering states with a fixed third component of the total 
angular momentum with respect to the proton momentum, $J_z$, and with a spin 
component $s =\pm 1/2$ of the quark along the unit vector {\bf s} introduced 
in the previous section. For the sake of simplicity, I assume $X$ to have 
spin zero, moreover I choose a state with $J_z$ = 1/2. Then 
\begin{equation}
|J_z=1/2; s; X\rangle = \alpha |\rightarrow, L_z=0; s; X\rangle + \beta 
|\leftarrow, L_z=1; s; X\rangle.
\end{equation}
Here $\rightarrow(\leftarrow)$ and $L_z$ denote the components along 
${\bf n}$, respectively, of the quark spin and orbital angular momentum, while
$\alpha$ and $\beta$ are Clebsch-Gordan coefficients. 
Then the probability of finding a quark with $J_z$ = 1/2 and spin component $s$ 
along ${\bf s}$, in a longitudinally polarized proton with a positive helicity, 
is
\begin{eqnarray}
|\langle P,\Lambda =1/2|J_z=1/2; s; X\rangle|^2 &=& \alpha^2 
|\langle P,\Lambda =1/2|\rightarrow, L_z=0; s; X\rangle|^2 \nonumber
\\
&+& \beta^2 |\langle P,\Lambda =1/2|\leftarrow, L_z=1; s; X\rangle|^2 + I, 
\label{prob}
\end{eqnarray}
\begin{equation}
I = 2 \alpha \beta Re\left[\langle P,\Lambda =1/2|\rightarrow, L_z=0; s; 
X\rangle 
\langle (\leftarrow, L_z=1; s; X)|P,\Lambda =1/2 \rangle\right].\label{intf} 
\end{equation}
Expanding the amplitudes in partial waves yields
\begin{equation}
I = 2 \sum_{l,l'=0}^{\infty} Re\left[ie^{-i\phi} A_l B^*_{l'}\right]
P_l(cos\theta) P^1_{l'}(cos\theta). \label{inter}
\end{equation}
Here $A_l$ and $B_l$ are related to partial wave amplitudes; moreover
$\theta$ and $\phi$ are respectively the polar and the azimuthal angle of 
the quark momentum, assuming ${\bf n}$ as the polar axis and, as the azimuthal 
plane, the one through ${\bf n}$ and ${\bf s}$. In the Breit frame one has
\begin{equation}
P_l(cos\theta) \sim 1, ~~~~~~ \ ~~~~~ P^1_l(cos\theta) \sim 
\frac{|{\bf p}_{1\perp}|}{x{\cal P}}.
\end{equation}
Then eq. (\ref{inter}) yields
\begin{equation}
I \sim \frac{|{\bf p}_{1\perp}|}{x{\cal P}}\left(A cos\phi + B sin\phi\right),
\label{intff}
\end{equation}
where $A$ and $B$ are real functions made up with $A_l$ and $B_l$. Since ${\bf
s}$ is an axial vector, parity conservation implies $A$ = 0. Therefore eqs. 
(\ref{prob}) and (\ref{intff}) imply that the interference term $I$ is T-odd
and that the final quark is polarized perpendicularly to the proton momentum 
and to the quark momentum, independent of the proton polarization. Comparing 
eq. (\ref{intff}) with eq. (\ref{polr}) yields 
\begin{equation}
\mu = x{\cal P}. \label{norm}
\end{equation}
Eqs. (\ref{norm}) predicts 
that the quark tranverse polarization in an 
unpolarized (or spinless) hadron decreases as ${\cal P}^{-1}$.

\section{Extracting chiral odd functions from DY}
\subsection{The transversity function}
Eqs. (\ref{diff0}), (\ref{tens}) and (\ref{chir}) imply that the numerator of 
the DY asymmetry (\ref{asy0}) is of the form
\begin{equation}
d\sigma_{\uparrow}-d\sigma_{\downarrow} \propto 
\int d^2p_{1\perp} \left[\left(\frac{p_{2a}}{x_2{\cal P}} h_{1T} + \frac{{\bf 
p_1}_{\perp}\cdot{\bf p_2}_{\perp}}{x_1x_2{\cal P}^2}h_1^{\perp} \right)
{\bar h}_1^{\perp} + (1\leftrightarrow 2)\right],\label{conv}
\end{equation}
assuming the constraint (\ref{cons}). Here 
\begin{equation}
p_{2a} = {\bf p}_{2\perp}\cdot {\bf S}\times{\bf n}.
\end{equation}
In order to extract the transversity, {\it i. e.}, $h_1$ = $\int 
d^2p_{\perp}h_{1T}$, from DY, I define the following weighted asymmetry[10,11]:
\begin{equation}
\langle A_1\rangle = \frac{\sum_nd\sigma^{(n)} Q_a^{(n)}}{M_P
\sum_nd\sigma^{(n)}},
~~~~~~~ \ ~~~~~~ Q_a = p_{1a} + p_{2a}. \label{asy}
\end{equation}
Here $M_P$ is the proton rest mass and $d\sigma^{(n)}$ the differential cross 
section at a fixed transverse momentum, the sum running over the data. Eq. 
(\ref{conv}) implies
\begin{eqnarray}
\sum_nd\sigma^{(n)} Q_a^{(n)} &\propto& h_1(x_1) 
{\bar h}_{1(1)}^{\perp}(x_2) + {\bar h}_1(x_1) h_{1(1)}^{\perp}(x_2), \label{we}
\\ 
h_{1(1)}^{\perp}(x_2) &=& \int d^2p_{\perp} p_{2a}^2h_1^{\perp}. \label{h11}
\end{eqnarray}
This allows to extract $h_1$ and ${\bar h}_1$, provided $h_{1(1)}^{\perp}$ and
${\bar h}_{1(1)}^{\perp}$ are 
known. These functions have to be inferred from an independent analysis, for 
example with the method I exhibit in the next subsection. According to my model,
one has $\langle A_1\rangle \propto {\cal P}^{-1}$.

\subsection{The function $h_1^{\perp}$}

$h_1^{\perp}$, which is washed out by the weighted asymmetry (\ref{asy}), can 
be singled out by using an alternative weight function. Indeed, defining 
\begin{equation}
\langle A'_1\rangle = \frac{\sum_n(d\sigma_{\uparrow}^{(n)}-
d\sigma_{\downarrow}^{(n)})(Q_a^{(n)})^2}{M^2_P\sum_nd\sigma^{(n)}},
\label{asy1}
\end{equation}
formula (\ref{conv}) yields
\begin{equation}
\langle A'_1\rangle \propto h_{1(1)}^{\perp}(x_1) {\bar h}_{1(1)}^{\perp}(x_2)
+ {\bar h}_{1(1)}^{\perp}(x_1) h_{1(1)}^{\perp}(x_2),
\label{we1}
\end{equation}
where $h_{1(1)}^{\perp}(x_1)$ is defined analogously to eq. (\ref{h11}).
Formula (\ref{conv}) implies that the weight functions ${\bf Q}^2_{\perp}$ 
and $({\bf Q}_{\perp}\cdot{\bf S})^2$ could be used instead of $Q_a^2$.
Moreover $h_1^{\perp}$ may be extracted also from unpolarized DY; in 
this case the weighted asymmetry is defined as
\begin{equation}
\langle A''_1\rangle = \frac{\sum_{n_+}d\sigma^{(n_+)}\left[{\bf 
Q}_{\perp}^{(n+)}\right]^2-\sum_{n_-}d\sigma^{(n_-)}\left[{\bf 
Q}_{\perp}^{(n_-)}\right]^2} {M^2_P\sum_nd\sigma^{(n)}}. \label{asy11}
\end{equation}
Here the symbols $\pm$ refer to events whose transverse momenta ${\bf 
Q}_{\perp}$ are at the right (+) or at the left (-) of the plane through 
${\bf n}$ and the unit vector ${\bf s}$. The model I have elaborated predicts 
that both $\langle A'_1\rangle$ and $\langle A''_1\rangle$ decrease as 
${\cal P}^{-2}$. 

\section{Discussion}
I have suggested how to extract $h_1$ and $h_1^{\perp}$ from DY. In particular, 
the T-odd function $h_1^{\perp}$ can be inferred from unpolarized proton-proton
collisions and it can be used as a quark polarimeter in order to get $h_1$ from 
single transverse spin asymmetry. Similar methods could be elaborated for 
semi-inclusive deep inelastic scattering(SIDIS)[12], for $e^+e^-$ $\to$ $\pi X$ 
and, with some assumptions, also for $p p$ $\to$ $\pi X$. In these reactions, 
in order to infer the tranversity, the Collins function[13] is needed, which 
can be deduced, for example, from $e^+e^-$ collisions.
Moreover in such kinds of experiments, as well as in DY, one is 
faced with the problem of disentangling the contributions of quarks and 
antiquarks of any flavor, as pointed out by Boglione and Leader[14]. In this 
connection, a comparison between SIDIS and DY results is particularly 
helpful, since the two cross sections depend on the quark and antiquark 
functions according to different combinations.


\begin{thebibliography}{99}

\bibitem{a} J. Collins: Nucl. Phys. B {\bf 396} (1993) 161 

\bibitem{b} R.L. Jaffe: Phil. Trans. Roy. Soc. Lond. A {\bf 359} (2001) 391

\bibitem{c} A. V. Efremov: hep-ph/0001214 and hep-ph/0101057

\bibitem{d} X. Ji: Phys Lett. B {\bf 284} (1992) 137

\bibitem{e} R.L. Jaffe, X. Jin and J. Tang: Phys. Rev. Lett. {\bf 80} (1998) 
1166; Phys. Rev. D {\bf 57} (1998) 5920

\bibitem{f} R.L Jaffe: "2nd Topical Workshop on Deep Inelastic Scattering off
polarized targets: Theory meets Experiment (Spin 97)", Proceedings eds. J. 
Bl\"umlein, A.  De Roeck, T. Gehrmann and W.-D. Nowak. Zeuten, DESY, (1997) p.
167 

\bibitem{g} D. Boer and P.J. Mulders: Phys. Rev. D {\bf 57} (1998) 5780

\bibitem{h} R.J. Ralston and D.E. Soper: Nucl Phys. B {\bf 152} (1979) 109

\bibitem{i} P.J. Mulders and R.D. Tangerman: Nucl Phys. B {\bf 461} (1996) 197

\bibitem{l} A.M. Kotzinian and P.J. Mulders: Phys. Rev. D {\bf 54} (1996) 1229

\bibitem{m} A.M. Kotzinian and P.J. Mulders: Phys. Lett. B {\bf 406} (1997) 
373

\bibitem{n} HERMES coll., Airapetian et al.: Phys. Rev. Lett. {\bf 84} (2000) 
4047 

\bibitem{o} J. Collins: Nucl Phys. B {\bf 396} (1993) 161

\bibitem{p} M. Boglione and E. Leader: Phys. Rev. D {\bf 61} (2000) 114001

\end{thebibliography}
\end{document}